\documentclass[twocolumn,aps,showpacs,floatfix,amsmath,amssymb,superscriptaddress,prb]{revtex4}

\usepackage{graphicx}
\usepackage{bm}

\newcommand{\up}{\uparrow}
\newcommand{\dn}{\downarrow}
\newcommand{\bc}{\begin{center}}
\newcommand{\ec}{\end{center}}
\newcommand{\be}{\begin{equation}}
\newcommand{\ee}{\end{equation}}
\newcommand{\ba}{\begin{array}}
\newcommand{\ea}{\end{array}}
\newcommand{\beq}{\begin{eqnarray}}
\newcommand{\eeq}{\end{eqnarray}}
\newcommand{\ket}[1]{\left| {#1}\right\rangle}
\newcommand{\bra}[1]{\left\langle {#1} \right|}
\newcommand{\skp}[2]{\langle {#1} | {#2}\rangle}
\newcommand{\e}[1]{\langle {#1}\rangle}
\newcommand{\trip}[3]{\langle {#1}|{#2}|{#3}\rangle}

\begin{document}

\title{Definitions of entanglement entropy of spin systems in the valence-bond basis}

\author{Yu-Cheng Lin}
\affiliation{Graduate Institute of Applied Physics, National Chengchi University, Taipei, Taiwan}

\author{Anders W. Sandvik}
\affiliation{Department of Physics, Boston University, 590 Commonwealth Avenue, Boston, Massachusetts 02215, USA}

\date{\today}

\begin{abstract}
The valence-bond structure of spin-1/2 Heisenberg antiferromagnets is closely related to quantum entanglement. 
We investigate measures of entanglement entropy based on transition graphs, which characterize state overlaps in
the overcomplete valence-bond basis. The transition graphs can be generated using projector Monte Carlo simulations 
of ground states of specific hamiltonians or using importance-sampling of valence-bond configurations of 
amplitude-product states. We consider definitions of entanglement entropy based on the bonds or loops shared by two
subsystems (bipartite entanglement). Results for the bond-based definition agrees with a previously studied definition 
using valence-bond wave functions (instead of the transition graphs, which involve two states). For the one dimensional 
Heisenberg chain, with uniform or random coupling constants, the prefactor of the logarithmic divergence with the size of 
the smaller subsystem agrees with exact results. For the ground state of the two-dimensional Heisenberg model (and also 
N\'eel-ordered amplitude-product states), there is a similar multiplicative violation of the area law. In contrast, 
the loop-based entropy obeys the area law in two dimensions, while still violating it in one dimension---both behaviors 
in accord with expectations for proper measures of entanglement entropy.

\end{abstract}


\maketitle
\section{Introduction}
The concept of quantum mechanical entanglement is of broad interest in physics.
One widely used quantitative measure of entanglement is the von Neumann
entanglement entropy $S_{\rm vN}$. Partitioning a many-body system in a pure
quantum state into two contiguous subsystems, $S_{\rm vN}$ is defined as the
von Neumann entropy of the reduced density matrix of either one of the two
subsystems. The insight that $S_{\rm vN}$ typically scales proportionally to
the boundary area (the area law \cite{EISERT}) was first developed in the context 
of black holes.\cite{SREDN} The area law has now become a key benchmark for
characterizing states of interest in quantum information theory \cite{KITAEV,WOLF} 
and condensed matter physics.\cite{LEVIN,FRADKIN} An important application of this 
concept is to construct computationally tractable variational ans\"atze for ground 
states based on the area law.\cite{TENSOR,TAGLIA}

While $S_{\rm vN}$ is a clear and well established measure of entanglement, it is difficult to compute 
for strongly-correlated quantum systems in dimensions higher than one, where exact diagonalization 
and density matrix renormalization group approaches become
inefficient.\cite{DMRG,LADDER}  Alternative definitions of entanglement
entropy (or, more precisely, alternative measures of entanglement) are therefore 
also actively investigated.  The R\'enyi entropies $S_n$ ($n=1,2,\ldots$) are often 
used,\cite{LI} and have the appealing property that $S_1=S_{\rm vN}$.  Recently it was 
realized \cite{RENYI} that $S_2$ (and in principle also $S_n$ for $n>2$) can be 
computed for quantum spin systems using projector quantum Monte Carlo (QMC) simulations 
in the valence bond (VB) basis.\cite{SANDVIK} Previously, a measure $S_{\text{VB}}$ of 
entanglement entropy explicitly making use of the VB basis was also introduced \cite{VBE1,VBE2} 
within the context of this QMC method. Generalizing an exact result for a single 
VB state \cite{REFAEL} to a superposition, $S_{\text{VB}}$ is 
given by the average number of VBs connecting two subsystems. While defined explicitly 
using the VB basis, this quantity also can be evaluated using the density matrix 
renormalization group (DMRG) method \cite{CAPPONI} and is, therefore, not completely 
tied to the VB basis. Its asymptotic behavior has also been found exactly using analytical 
methods for the Heisenberg chain.\cite{JACOBSEN}

In this paper, we formulate a different measure $S_{\rm VB}^2$ of entanglement entropy for quantum spin systems 
in terms of the transition graphs characterizing overlaps of VB basis states. The transition graphs are 
generated in projector Monte Carlo simulations of the ground state of a hamiltonian,\cite{SANDVIK,LOOP} 
or in Monte Carlo sampling of bond configurations of variational states such as the amplitude product 
states.\cite{LIANG1,LOOP} Like the previous definition of VB entanglement entropy, which we henceforth call 
$S_{\text{VB}}^1$, the transition-graph definition $S_{\rm VB}^2$ involves VBs shared by two subsystems, 
but the weighting is different because the bond configurations in the projected bra and ket states 
are sampled using their individual wave functions and overlap (which depends on the number of loops in the 
transition graph).  We show that $S_{\rm VB}^2$ scales with the subsystem size in accord with an exact result \cite{JACOBSEN} 
for the previous definition $S_{\rm VB}^1$ in the one dimensional (1D) Heisenberg chain. The 
corrections to this form are much smaller than in the previous definition, however. Thus, the 
different weighting procedure appears to reduce the subleading size corrections. 

Both $S_{\rm VB}^1$ and $S_{\rm VB}^2$ violate the area law in the case of the N\'eel-ordered ground state 
of the two-dimensional (2D) Heisenberg model. We here argue that this is because definitions based on 
single VBs typically will overestimate the entanglement, due to the over-completeness of the VB basis. To 
remedy this, we propose  an alternative measure $S_{\rm loop}$ of entanglement entropy based on the loops 
of the transition graphs, i.e., $S_{\rm loop}$ is the average number of transition-graph loops shared by the 
two subsystems. Loops correspond to maximally entangled groups of spins, and the number of loops 
shared by the subsystems is therefore an appropriate measure of entanglement. We show that $S_{\rm loop}$ 
of the 2D Heisenberg model (and also in a generic variational amplitude-product state with N\'eel
order) obeys the area law and has an additive logarithmic correction, in contrast to the multiplicative 
logarithmic corrections affecting the bond-based definitions. Thus, it appears that the loop entropy scales 
in the same way as the R\'enyi entropy (as observed in Ref.~\onlinecite{RENYI}) and the  standard von Neumann 
entanglement entropy (where one would expect such scaling) and, thus, may be a convenient (more easily 
computable) stand-in for these definitions.

The outline of the rest of the paper is as follows. In Sec.~\ref{vbmethods} we summarize the properties 
of the VB basis needed for our definitions and calculations, and also briefly review amplitude-product 
states,\cite{LIANG1} the VB projector QMC method,\cite{SANDVIK,LOOP} and the loop-gas picture \cite{SUTHER1,SUTHER2} 
of VB states. The VB and loop entropy definitions are discussed and tested in Secs.~\ref{bondentropy} and 
\ref{loopentropy}. We conclude in Sec.~\ref{summary} with a summary and discussion.

\section{Valence-bond basis and methods}
\label{vbmethods}

We will consider the spin-$1/2$ Heisenberg hamiltonian, written in the form
\be
     H=-\sum_{\e{i,j}} J_{ij} S_{ij},~~~~~(J_{ij}>0),\label{eq:h}
\ee
where $\e{i,j}$ denotes nearest-neighbor spins on a lattice with periodic boundaries and
$S_{ij}$ is a singlet projector,
\be
S_{ij}=1/4-\mathbf{S}_i\cdot\mathbf{S}_j,
\ee
and $J_{ij}>0$ are antiferromagnetic coupling constants. We will study 1D chains with uniform and random couplings, as well 
as uniform 2D square lattices. 

Here, in Sec.~\ref{basis}, we discuss the VB basis in which we carry out all calculations. In Sec.~\ref{aprodstates} we discuss variational VB
amplitude-product states, which give some important insights into various types of ground states and their entanglement properties. The projector 
QMC calculations that we use for unbiased calculations are briefly reviewed in Sec.~\ref{qmcmethod}. The loop-gas picture of VB states was 
first introduced by Sutherland.\cite{SUTHER1,SUTHER2} In Sec.~\ref{loopgas} we discuss it in a somewhat broader sense, which we will rely on 
for the definition of loop entropy in Sec.~\ref{loopentropy}.

\subsection{The valence-bond basis}
\label{basis}

The ground state of $H$ on a bipartite lattice with an even number $N$ of spins is a total-spin singlet
and can be expanded in bipartite VB states (i.e., each bond connects sites on the two sublattices) with all positive 
coefficients;\cite{LIANG1}
\be
\ket{\psi_0}=\sum_v \lambda_v\ket{v},~~~~(\lambda_v\ge 0\; \forall~ v).
\label{eq:vexp}
\ee
With the singlet state of two spins $i,j$ denoted by $(i,j)$, a bipartite VB basis state is defined as
\be
\ket{v}:=\bigotimes_{ij}(i,j),~~~~
(i,j):=\frac{1}{\sqrt{2}}\bigl ( \ket{\up_i\dn_j}-\ket{\dn_i\up_j} \bigr ),
\label{singij}
\ee
where sites $i$ and $j$ are on different sublattices and each site appears exactly once in the product.
Thus, there are $(N/2)!$ different basis states $\ket{v}$ which form an over-complete basis in the singlet subspace. 

\begin{figure}
\includegraphics[width=8cm, clip]{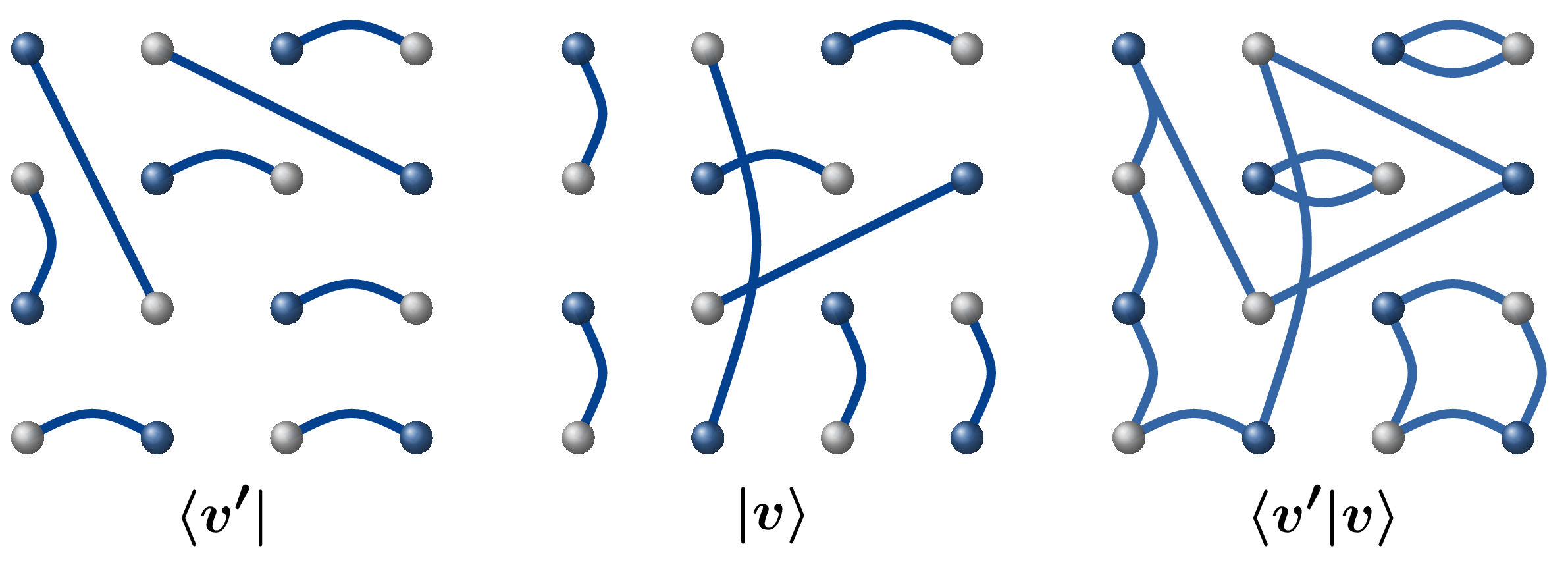}
\caption{
    \label{fig:loop}
   (Color online) Transition graph (right) determining the overlap of two VB states $\ket{v}$ and $\ket{v'}$. Note that in the 
   restricted VB basis considered here, all bonds connect sites on different sublattices (with sites on sublattices $\mathcal{A}$ 
   and $\mathcal{B}$ here indicated by darker and lighter circles). The cluster of spins defined by a loop can be in two states,
   with all spins $\up$ on sublattice $\mathcal{A}$ and $\dn$ on $\mathcal{B}$, or vice versa.} 
   \end{figure}

The overlap $\skp{v'}{v}$ between two VB basis states can be expressed in terms of transition-graph loops, 
as illustrated in Fig.~\ref{fig:loop}. Each lattice site is connected to one bond from $\ket{v}$ and one from $\bra{v'}$, 
and all the bonds therefore form closed loops. The matrix element $\skp{v'}{v}$ is a product of factors arising from 
these loops. To evaluate $\skp{v'}{v}$ it is more convenient to work in the standard spin-$z$ basis and rewrite a VB state $\ket{v}$ as
\be
  \ket{v}=\frac{1}{2^{N/4}}\sum_{\alpha=1}^{2^{N/2}} (-1)^{n_{\mathcal{B}\up}} \ket{S^z_1,S^z_2,\cdots,S^z_N}_\alpha,
\label{vbszpsi}
\ee
where $\alpha$ labels the spin states that are compatible with the valence bonds in $|v\rangle$ (i.e., $\up\dn$ or $\dn\up$ spin configurations 
on each bond) and $n_{\mathcal{B}\uparrow}$ is the number of $\up$ spins on sublattice $\mathcal{B}$, the sign following if the singlet (\ref{singij}) is 
defined such that $i$ and $j$ are on sublattices $\mathcal{A}$ and $\mathcal{B}$, respectively (which corresponds to Marshall's sign rule \cite{MARSH} 
for a bipartite system). From the orthonormality of the ordinary basis of states $\{\ket{S^z_1,S^z_2,\cdots,S^z_N}\}$, the only nonzero terms 
in $\skp{v'}{v}$ expressed using (\ref{vbszpsi}) are those with spin configurations common to both the VB states $\ket{v}$ and $\ket{v'}$. 
This corresponds to spins forming staggered patterns, $\up\dn\up\cdots \dn$ or $\dn\up\dn \cdots \up$, around each loop. There are two such staggered 
configurations of each loop and the signs in (\ref{vbszpsi}) cancel in the overlap, which, thus, is given by\cite{SUTHER1}
\be
      \skp{v'}{v}=2^{N_\circ-N/2},
\label{wloop}
\ee
where $N_\circ$ is the number of loops in the transition graph. This is illustrated in Fig.~\ref{fig:loop}.

In the VB description many physically quantities are determined by the statistical properties of the transition-graph loops. 
For instance, the normalized matrix element needed for computing the spin-spin correlation function is given by 
\be
\frac{\langle v'|{\bf S}_i \cdot {\bf S}_j|v\rangle}{\langle v'|v\rangle} = \left \lbrace
\begin{array}{rl}
\pm 3/4, & ~~[i,j], \\
0, & ~~[i][j], \\
\end{array}\right.
\label{spincorr}
\ee
where $[i,j]$ and $[i][j]$ denote sites $i$ and $j$ belonging to the same loop and different loops, respectively,
and the sign in the case $[i,j]$ is $+$ and $-$ for spins on the same and different sublattices, respectively.

Note that the transition graph loops define clusters of spins that do not necessarily resemble geometric loops on the lattice, 
because the VBs can be of any length and the bonds forming a loops can cross each other (as seen in Fig.~\ref{fig:loop}). Note also that the 
staggered spin configuration along a loop also corresponds to staggering of all spins in the loop in the sense of the two sublattices, 
i.e., all spins on sublattice $\mathcal{A}$ are parallel and opposite to those on $\mathcal{B}$. This is the origin of the signs in (\ref{spincorr}).

\subsection{Amplitude-product states}
\label{aprodstates}

In an amplitude-product state,\cite{LIANG1} the expansion coefficients in (\ref{eq:vexp}) are products 
of amplitudes $h({\bf r}_{ij})>0$ (to maintain Marshall's sign rule for the ground state of a bipartite model),
\be
\lambda_{{v}}=\prod_{(i,j)} {h({\bf r}_{ij})},
\label{lambdaprod}
\ee
with the vector ${\bf r}_{ij}$ denoting the ``shape'' (the lengths in all lattice dimensions) of bond $(i,j)$ in 
$\ket{{v}}$. This form applies to a  translationally invariant system, while in a non-uniform system one should use 
$h({\bf r}_{i},{\bf r}_{j})$ depending on both end-points of the bonds. 

For a given set of amplitudes, the expectation value of some operator $O$ can be written as
\be
       \e{O}=\frac{\trip{\psi_0}{O}{\psi_0}}{\skp{\psi_0}{\psi_0}}=
         \frac{\sum_{v v'} W_{v v'} \frac{\trip{v'}{O}{v}}{\skp{v'}{v}}}{\sum_{v v'} W_{v v'}},
        \label{eq:o}
\ee  
where the configuration weight is given by
\be
W_{v v'}=\lambda_v \lambda_{v'}\,\skp{v'}{v} = \lambda_v \lambda_{v'}2^{N_\circ-N/2}.
\label{Wvv}
\ee
The ratio $\trip{v'}{O}{v}/\skp{v'}{v}$ can normally be related to the loop structure of the transition 
graph,\cite{LIANG1,BEACH3} e.g., Eq.~(\ref{spincorr}) in the case of a spin correlation function. 

The expectation value (\ref{eq:o}) is ideally suited for evaluation using Monte Carlo sampling methods,\cite{LIANG1,LOOP} and all the 
amplitudes can be variationally optimized.\cite{LOU,LOOP} Calculations for the N\'eel-ordered ground state of the 2D Heisenberg model suggest 
that the fully optimized amplitudes decay as a power-law, $h(|{\bf r}_{ij}|)=1/|{\bf r}_{ij}|^p$, with $p\approx 3$.\cite{LOOP,LOU} 
There is reason to believe that the decay exponent in fact is exactly $p=3$, as this is the exponent obtained in an analytical 
mean-field-like treatment.\cite{BEACH1} Some aspects of the critical 1D chain can also be captured with amplitude-product 
states.\cite{BEACH2}

The properties of quantum systems are often modified dramatically by introducing quenched (static) randomness: e.g., 
quantum phase transitions with disorder can lead to new universality classes. For the Heisenberg chain, it is known that any amount 
of quenched randomness will drive the system into a state well approximated by a random singlet state---a single VB basis state 
(which is of the ``nested'' type, with no crossing VBs) with arbitrary VB lengths obtained according to a strong-disorder 
renormalization-group (SDRG) scheme.\cite{DASGUP,FISHER,SDRG} In this case, the transition-graph loops coincide exactly with the 
VBs, i.e., $N_\circ=N/2$ in (\ref{wloop}), and many asymptotic properties follow directly from the length distribution of 
the VBs.

Amplitude-product states (as well as the special case of the SDRG random singlet states) are useful variational states and we will 
consider some aspects of their entanglement properties. In many cases completely unbiased results are needed, however. One way to achieve 
this is with projector QMC simulations in the VB basis, which we briefly discuss next.

\subsection{Projector QMC method}
\label{qmcmethod}

In the VB QMC method \cite{SANDVIK,LIANG2} the ground state of the hamiltonian
(\ref{eq:h}) is projected out stochastically, by applying a high power of $H$
to some trial singlet state $\ket{\psi_t}$ in the VB basis; $(-H)^m\ket{\psi_t}
\to \ket{\psi_0}$ for a large $m$ (up to an irrelevant normalization). A good
trial state, such as an optimized amplitude-product state for a uniform system or a single
VB state obtained with the SDRG procedure for a 1D random chain, can be used to
optimize the convergence properties of such a scheme, but the final result is not 
sensitive to $\ket{\psi_t}$ as long as $m$ is sufficiently large (i.e., the method is
unbiased). 

There are two formulations of the VB QMC method,\cite{SANDVIK} generating either the ground state wave function or the 
ket and bra versions of the ground state needed to evaluate expectation values. In the former case, stochastic application 
of the projector $H^m$ (for sufficiently large $m$) on the trial state produces VB basis states distributed proportionally 
to the expansion coefficients $\lambda_v$ in (\ref{eq:vexp}), i.e., these coefficients are not known (and would be much more 
complicated than the simple amplitude products) but importance-sampling of the contributions to 
$H^m|\psi_t\rangle$, which resemble terms in a path integral, generate them probabilistically 
(with many paths contributing to a single coefficient $\lambda_v$).

In the ``double projection'' method, an expectation value is formally given by Eq.~(\ref{eq:o}), but the expansion coefficients 
$\lambda_v$ are again not known. The sampling of paths obtained from the combined ket $H^m \ket{\psi_t}$ and  bra $\bra{\psi_t}H^m$ 
states, i.e., $\bra{\psi_t}H^{2m}\ket{\psi_t}$, leads to a series of VB-pair configurations distributed according to 
the weights $W_{v v'}$ in (\ref{Wvv}). 

In both the wave-function and expectation-value projection schemes, one can employ efficient loop updates for generating the states
and transition graphs, as discussed in detail in Ref.~\onlinecite{LOOP}. In the sampling procedures of such algorithms, one uses the combined 
spin-bond basis and applies a high power ${H}^{m}$ or ${H}^{2m}$ in a way similar to the ''operator-loop'' update in the finite temperature stochastic 
series expansion QMC method.\cite{SSE} The computational effort scales as $\mathcal{O}(m)$, and to converge calculated quantities to
the ground state $m$ has to be scaled as $\propto N^a$,  with $a$ typically in the range $1-2$ (depending on the model's low-energy energy 
spectrum and the quality of the trial state $|\psi_t\rangle$).\cite{LOOP}

\subsection{Loop-gas description}
\label{loopgas}

\begin{figure}
\includegraphics[width=5.75cm, clip]{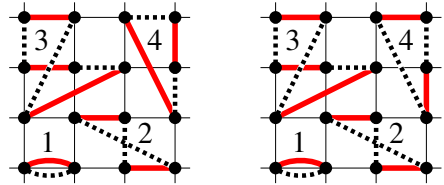}
\caption{
 \label{fig:loops}
  (Color online) Two transition graphs $\langle v'|v\rangle$ with the same loops structure (with four loops, labeled 1,2,3,4). The two 
  cases shown here illustrate how bonds within a loop (here loop 4) can be swapped between $|v\rangle$ and $\langle v'|$  (represented by solid and  
  dashed bonds, respectively). This kind of bond reconfiguration leaves the weight unchanged in the case of an amplitude-product state, but not in a more 
  general state in which there are correlations between the bonds.}
\end{figure}

The loop-gas view of VB sampling was suggested by Sutherland some time ago.\cite{SUTHER1,SUTHER2} Consider the two VB configurations 
in a transition graph $\langle v'|v\rangle$ with the associated weight (\ref{Wvv}) for an amplitude-product state. In that case, 
the factor $\lambda_v\lambda_{v'}$ does not depend on details of the bond arrangements, only on the total number $N_b({\bf r})$ of bonds 
of all shapes ${\bf r}$ in the two states $|v\rangle$ and $|v'\rangle$ (i.e., no bond correlations are included). In addition to the factor 
$2$ in the overlap (\ref{wloop}) for each loop in the transition graph, for each loop containing more than two bonds we can consider swapping 
all bonds belonging to a given loop between $|v\rangle$ and $|v'\rangle$, as illustrated in Fig.~\ref{fig:loops}. This swapping does not affect 
the weight (\ref{Wvv}) of the joint configuration of the two states (although the weights $\lambda_v$ and $\lambda_{v'}$ of the individual states 
are affected). The two bond configurations and two staggered spin patterns for each loop corresponds to a loop fugacity $4$ for all loops of 
length $4$ or larger. The shortest, length-$2$, loops have fugacity $2$, as the bond swapping in this case does not affect the bond configurations. 

Instead of sampling bonds weighted according to (\ref{Wvv}), one can think of sampling loops with the weight
\be
W=2^{N_\circ(2)}4^{N_\circ(>2)}\prod_{\bf r} h({\bf r})^{N_b({\bf r})},
\label{wloopgas}
\ee
where $N_\circ(2)$ and $N_\circ(>2)$ denote the number of loops of length $2$ and larger than $2$, respectively, and the unimportant factor $2^{-N/2}$ 
in (\ref{wloop}) has been omitted. The product of amplitudes can here be thought of as originating from the shapes of the loops. In special cases, 
such as Anderson's resonating valence-bond (RVB) state including only the shortest bonds (of length $r=1$),\cite{RVB,ALBU,TANG} the weight 
only depends on the number of loops. This was the case considered by Sutherland,\cite{SUTHER2} who also studied generalizations of the loop
gas in which the loop fugacities, $Z_2=2$ and $Z_{>2}=4$ in (\ref{wloopgas}), can take arbitrary values and, thus, phase transitions can be 
studied as a function of these fugacities.

The loop-gas weight function (\ref{wloopgas}) does not apply to states beyond the amplitude-product description. In the exact ground
state of a given hamiltonian, one would in general expect bond correlations. The product of two VB expansion coefficients in (\ref{Wvv}) then
changes when an intra-loop bond swap of the type illustrated in Fig.~\ref{fig:loops} is carried out, thus invalidating the form (\ref{wloopgas}).
One can still, however, sum up the weights obtained as a result of all such bond swaps which leave the loop structure intact, and this way, 
in principle (but hardly in practice), write down a weight for a loop configurations which is more complicated than (\ref{wloopgas}). The loop 
configurations  generated in the QMC projector method represent a stochastic implementation of this more general loop-gas picture. Note that even 
in this generalization, each loop is associated with a factor of two arising from the two allowed staggered spin configurations of the cluster 
of sites defined by the loop.

\section{Valence-bond entanglement entropy}
\label{bondentropy}

A singlet state formed by two $S=1/2$ spins is maximally entangled and has the maximum value of the von Neumann 
bipartite entanglement entropy; $S_{\text{vN}}=1$ (measured in bits). The simplest case for computing $S_{\text{vN}}$ of a
many-body system is a single VB basis state $\ket{v}$. For a given bipartition, its entanglement entropy is just the 
number of singlets $n_{AB}$ connecting subsystems $\mathcal{A}$ and $\mathcal{B}$.\cite{REFAEL} It is not easy to compute $S_{\text{vN}}$
for a superposition of VB states, however. In Refs.~\onlinecite{VBE1} and \onlinecite{VBE2} a straight-forward generalization of the 
result for a single VB state was proposed as an alternative definition of entanglement entropy for an arbitrary superposition of
VB states, using the average number of bonds connecting the two subsystems. With the VB projector method for the wave function, 
one stochastically evaluates 
\be
S_{\text{VB}}^{1}=\frac{\sum_v \lambda_v n_{AB}}{\sum_v \lambda_v}.
\label{svb1def}
\ee
Here we investigate a different definition, using the expectation value 
\be
S_{\text{VB}}^{2}=\langle n_{\rm AB}\rangle,
\label{svb2def}
\ee
evaluated using Eq.~(\ref{eq:o}) with $O=n_{\rm AB}$. Here it should be noted that we define $n_{\rm AB}$ as just counting of the 
bonds crossing the two subsystems, and that this counting can be performed in the transition graph of $\langle v'|v\rangle$ either 
in $|v\rangle$ or $\langle v'|$. For a given configuration these numbers are different (i.e., the operator is not hermitian), but the 
averages are the same. In Sec.~\ref{summary} we will address further potential problems in interpreting $\langle n_{\rm AB}\rangle$ 
as a standard quantum mechanical expectation value. For now, we consider this quantity as an interesting aspect of the transition 
graphs, the statistical properties of which are uniquely determined for a given hamiltonian with short-range interactions (as we will 
further argue in Sec.~\ref{summary}).

For a single VB state $S_{\text{VB}}^{2}=S_{\text{VB}}^{1}$, but in a superposition these quantities are different, because of the over-completeness 
and associated different weighting of the VB configurations. We next present projector QMC results for both $S_{\text{VB}}^{1}$ and $S_{\text{VB}}^{2}$ 
in 1D and 2D Heisenberg systems and discuss their scaling as a function of the size of the smaller subsystem of the bipartition.

\subsection{Homogeneous chain}

\begin{figure}
\includegraphics[width=8cm, clip]{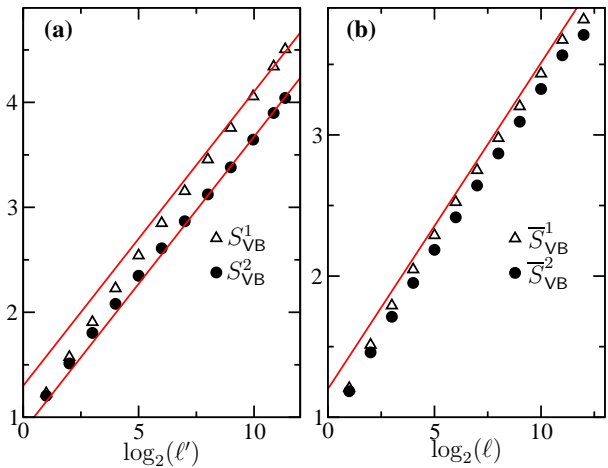}
\caption{
  \label{fig3}
  (Color online) (a) VB  entropies for subsystems of size $\ell$ in a Heisenberg chain of 
  length $L=2^{13}$ as a function of the logarithm of the conformal distance; 
$\log_2(\ell')=\log_2[(L/\pi)\sin(\pi\ell/L)]$.   The lines have slope $\gamma=4\ln(2)/\pi^2$, 
as in the exact asymptotic form.\cite{JACOBSEN} (b) Disorder-averaged VB entropies for the 
1D random chain, as a function of $\log_2(\ell)$. The line has slope $\gamma=\ln(2)/3$.} 
\end{figure}

We consider first the standard Heisenberg chain with uniform interactions. For
a large segment of size $\ell$, embedded in a chain of length $L$, the von
Neumann entanglement entropy has the asymptotic behavior
$S_{\text{vN}}(\ell)=(c/3)\log_2(\ell')+s_0$, where
$\ell'=(L/\pi)\sin(\pi\ell/L)$ is the conformal length, $c=1$ is the central
charge, and $s_0$ is a non-universal constant.\cite{CARDY} The VB entropy
$S^1_{\rm VB}$ is known to diverge in the same way, but with a different
factor, $\gamma=4\ln(2)/\pi^2 \approx 0.281$.\cite{JACOBSEN} Previous
calculations are consistent with $\gamma<1/3$ for large chains,\cite{LADDER}
but, as can be seen in Fig.~\ref{fig3}(a) for a chain of $2^{13}$ spins, there
are still large non-asymptotic corrections which make it difficult to verify
the factor precisely.\cite{CAPPONI} On the other hand, our results for
$S_{\text{VB}}^{2}$ are completely consistent with the known $\gamma$ over a
large range of subsystems. This may appear surprising, because the exact
calculation is based on the wave-function definition $S^1_{\rm
VB}$,\cite{JACOBSEN} not the expectation value $S^2_{\rm VB}$. It is plausible,
however, that $S^1_{\rm VB}$ and $S^2_{\rm VB}$ scale asymptotically with the
same $\gamma$ (with different additive constants), but that $S^2_{\rm VB}$ is
less affected  by subleading scaling corrections.

\subsection{Disordered chain}

Now we turn to the disordered chain, with random couplings generated from the
uniform distribution in the interval $(0,1]$. The ground state of this system
is known to be well approximated by a random singlet, where all spins form a
single non-crossing VB state with arbitrary bipartite bond lengths.\cite{FISHER} Using 
an SDRG analysis, Refael and Moore \cite{REFAEL} showed that the disorder average
of the von Neumann entanglement entropy $\overline{S}_{\text{vN}}$ in such a
state scales logarithmically with a universal coefficient $\gamma=\ln(2)/3$.
This result should hold also for the VB entanglement entropy of the Heisenberg
chain when $L\to\infty$, although the ground state is not exactly a single VB
basis state---there are fluctuations around the dominant SDRG configuration.\cite{TRAN} 
Our results for both $S^1_{\rm VB}$ and $S^2_{\rm VB}$, shown in Fig.~\ref{fig3}(b), 
are consistent with $\gamma=\ln(2)/3$ (considering some remaining finite-size effects 
for the largest subsystems). 

\subsection{Two-dimensional system}

\begin{figure}
\includegraphics[width=7.25cm, clip]{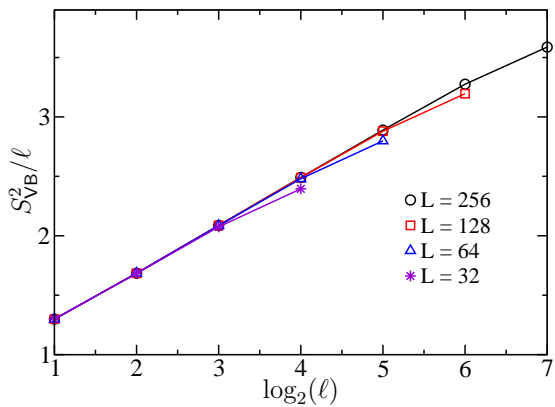}
\caption{
  \label{fig4}
  (Color online) The VB entropy $S_{\text{VB}}^2$ divided by the linear 
  dimension $\ell$ of a square sub-block in 2D $L\times L$ lattices with $L=32-256$.}
\end{figure}

We next consider the 2D Heisenberg model, which we have studied on $L\times L$ lattices 
with $L$ up to $256$. Fig.~\ref{fig4} shows results for $S_{\text{VB}}^2(\ell)$ of 
square subsystems of linear size $\ell$. The data converge very rapidly with $L$ for 
$\ell\le L/4$. There is clearly a multiplicative logarithmic correction to the area law, 
$S_{\text{VB}}^2(\ell)/\ell\propto \log_2(\ell)$, as found previously for $S_{\text{VB}}^{1}$ 
in Refs.~\onlinecite{VBE1,LADDER}. Both VB entropy definitions, thus, violate the area 
law in this case. The von Neumann entropy, on the other hand, should obey the area law, 
$S_{\text{vN}}(\ell)\propto \ell$, although, because of difficulties in calculating this 
quantity in an unbiased way, it has not been possible to confirm it unambiguously.\cite{LADDER} 
The recent calculation of the R\'enyi entropy $S_2$ is, however, in agreement with the area
law.\cite{RENYI}

We conclude that $S_{\text{VB}}^2$ has an advantage over $S_{\text{VB}}^1$, in the sense that its logarithmic 
prefactor in 1D converges faster to the result of the exact calculation in Ref.~\onlinecite{JACOBSEN} (although
we do not know whether the faster convergence is generic). Neither of these definitions serves well as a proxy
for the von Neumann entanglement entropy $S_{\text{vN}}$, however, because the area law is violated in 2D by a 
multiplicative logarithmic correction. Only additive corrections to the area law are expected.\cite{CASINI,RYU}

\section{Loop entanglement entropy}
\label{loopentropy}

The loop-gas picture discussed in Sec.~\ref{loopgas} suggests a potential reason why the VB entanglement entropies (\ref{svb1def}) and 
(\ref{svb2def}) violate the area law in 2D systems: The VB basis consists of singlet pairs, but when considering an expectation value, 
constraints related to the over-completeness are imposed on the spin states beyond the singlet pairing. These constraints are similar to 
multi-spin entanglement, as there are two spin states for each loop (the two staggered spin configurations on the loops) and these loops are, 
thus, analogous to maximally entangled sets of spins (albeit in an overlap matrix element, not a wave function). All bonds therefore do not 
carry a full unit of entanglement entropy (because they are entangled with other bonds in the same loop), and the VB entropy $S_{\text{VB}}^2$ 
may therefore overestimate the actual entanglement entropy. This could also be the case with $S_{\text{VB}}^1$, because it, too, is based 
on a superposition of states with only pair-wise entanglement (and, as we showed in Sec.~\ref{bondentropy}, $S_{\text{VB}}^1$ and $S_{\text{VB}}^2$ 
have the same scaling properties). On the other hand, in 1D $S_{\text{VB}}^1$ and $S_{\text{VB}}^2$ actually underestimate the entanglement 
entropy (in relation to the von Neumann entanglement entropy, which is slightly larger than the bond entropies in this case). The intuitive
picture of entanglement entropy in terms of bonds is therefore, as a consequence of the overcompleteness, not always quantitatively correct.

\begin{figure}
\includegraphics[width=3.25cm, clip]{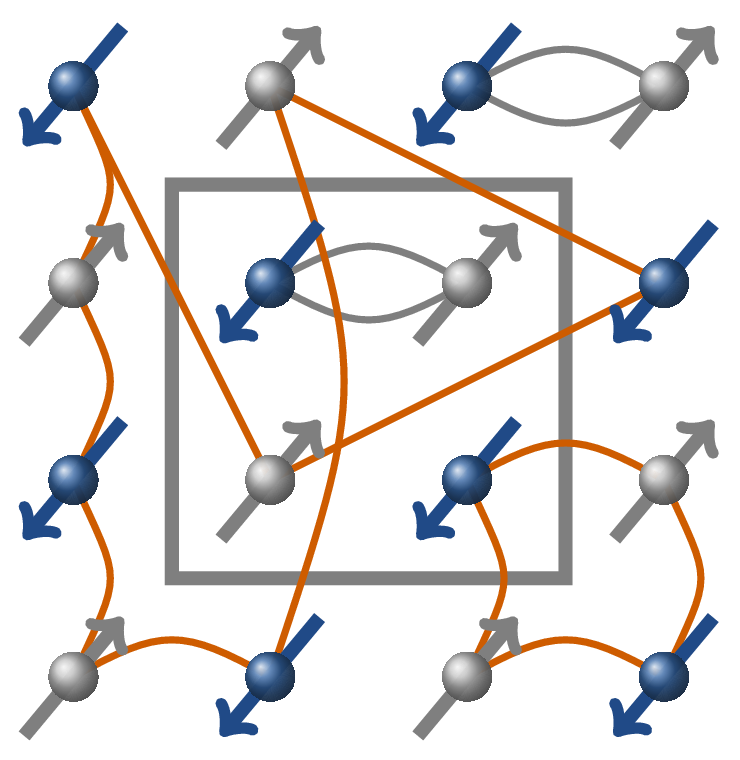}
 \caption{
  \label{fig5}
  (Color online) The loop entropy $S_{\text{loop}}$ of a subsystem (gray square) is defined as the average number of loops
  in the transition graph shared by the subsystem and the rest of the system. In this example two loops are shared.}
\end{figure}

We will here consider a measure of entanglement entropy $S_{\rm loop}$ in terms of shared loops in the transition graph, as illustrated 
in Fig.~\ref{fig5}. Since the number of shared loops must be smaller or equal to the number of shared bonds, we have $S_{\rm loop} \le S_{\text{VB}}^2$, 
and it is therefore clear that $S_{\rm loop}$ cannot be a better approximation to $S_{\rm vN}$ than  $S_{\text{VB}}^2$ in the case of the 1D Heisenberg 
chain. We will be mainly interested in $S_{\rm loop}$ of the 2D system, but will also investigate it in 1D.

Before discussing the actual definition of $S_{\rm loop}$ further, let us consider for a moment the wave function. 
Instead of a bond-singlet pairing of the spins, one might regard a state as a superposition of products of loop-cluster states,
\be
|\Psi\rangle = \sum_{\alpha}\bigotimes_{i=1}^{N_{\alpha}} |\psi(i)\rangle_\alpha,
\ee
where $N_\alpha$ is the number of loops in component $\alpha$ and an individual cluster state of $n$ spins is of the form
\be
 |\psi(i) \rangle_{\alpha} = \frac{1}{\sqrt{2}}\left(\ket{\up_1\dn_2\cdots \dn_n}_{\alpha}^i-
      \ket{\dn_1\up_2 \cdots \up_n}_{\alpha}^i \right).
\ee   
This is a maximally entangled state of two staggered spin configurations along the loop of $n$ spins (or, equivalently, all even numbered 
spins are on sublattice $\mathcal{A}$ and all odd ones on $\mathcal{B}$, or vice versa). With such as loop-cluster wave function, it is natural 
to regard a loop shared by two subsystems in a bipartition as carrying one unit of entanglement entropy. Thus,
we define the {\it loop entanglement entropy} as
\be
S_{\rm loop}=\langle \Lambda_{AB}\rangle,
\ee 
where $\Lambda_{AB}$ counts the number of loops shared by subsystems $A$ and $B$ (i.e., loops passing through both the subsystems) in a
given bipartition. 

The loop-cluster view of entanglement entropy is realized in SDRG calculations for the random transverse-field Ising model,\cite{LIN} where the 
clusters have parallel spins and the ground state is given by a single product of such cluster states. The generalization to a superposition of
cluster products has the same motivation as the generalization of the entanglement entropy of the single VB state appearing in the SDRG scheme
for the random Heisenberg chain \cite{REFAEL} to a superposition.\cite{VBE1,VBE2} In general, it is not known, however, how to write the wave 
function of a Heisenberg system as a superposition of cluster states (and clearly there is no unique way of doing so, as such a basis is
massively overcomplete). As we have seen, such loop-cluster superpositions do appear in the loop-gas picture, and it is then natural to define 
$S_{\rm loop}$ as above using the transition graph loops, as illustrated in Fig.~\ref{fig5}. We will explore this measure of entanglement here. 

One may again question whether an expectation value such as $\langle \Lambda_{AB}\rangle$ is a {\it bona fide} quantum mechanical expectation value. 
We will discuss this further in Sec.~\ref{summary} and here only consider it, like $S^2_{\rm VB}$, as a statistical 
property of the transposition graphs generated in the projector QMC scheme (or in sampling of an amplitude-product state, 
which we will also investigate).

Clearly, the loop entropy is a boundary property, and we can write either $S_{\rm loop}(A)$ or $S_{\rm loop}(B)$ for a bipartition $(A,B)$. It is 
easy to demonstrate the sub-additive property $S_{\rm loop}(A_1\cup A_2)\le S_{\rm loop}(A_1)+S_{\rm loop}(A_2)$ for two different bipartitions, 
$(A_1,B_1)$ and $(A_2,B_2)$. The essential properties of an entropy are thus satisfied. For a single VB configuration, loops and bonds coincide and, 
thus, $S_{\rm loop}=S^1_{\rm VB}=S^2_{\rm VB}=S_{\rm vN}$ in this case.

\begin{figure}
\includegraphics[width=7.5cm, clip]{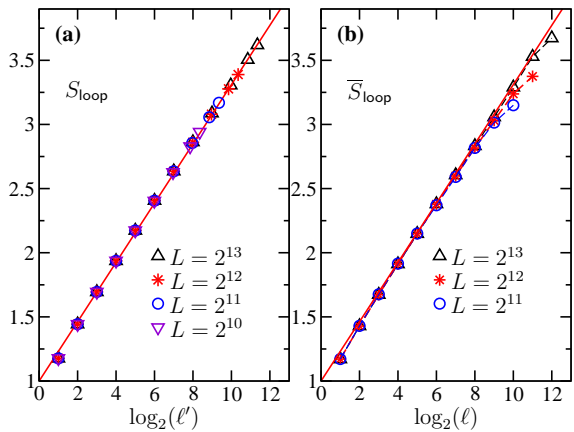}
\caption{
  \label{fig6}
  (Color online) Loop entropy of Heisenberg chains with (a) uniform and (b) random couplings (versus the
  logarithms of the conformal and linear subsystem sizes, respectively). Both lines have slope $\gamma=\ln(2)/3$.}
\end{figure}

\subsection{1D and 2D Heisenberg models}

\begin{figure}
\includegraphics[width=7.25cm, clip]{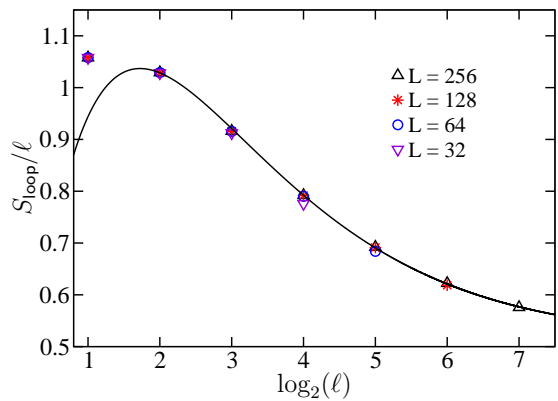}
\caption{
  \label{fig7}
  (Color online) The loop entropy divided by the linear dimension $\ell$ of a square sub-block in
  several $L\times L$ lattices. The curve shows a fit of the $L=256$ data ($\ell\ge 2^2$) to
  the form $S_{\text{loop}}=\mu \ell+a\log_2(\ell)+s_0$, with $\mu \approx 0.51$, $a \approx 1.12$, 
  and $s_0\approx -0.34$. The quality of this fit is statistically sound, with $\chi^2/{\rm dof}<1$ when 
  $\ell \ge 8$ data are included. The error bars are much smaller than the symbols, of the order $10^{-4}$ 
  for $S_{\rm loop}/\ell$.}
\end{figure}

We first consider the loop entropy for ground states of  Heisenberg models obtained by the projector QMC method.
Figs.~\ref{fig6} and \ref{fig7} show results for 1D and 2D systems, respectively. Interestingly, in 1D, the behavior is 
consistent with a logarithmic divergence with a prefactor $\gamma=\ln(2)/3\approx 0.231$ (to within a statistical
error of about 1\%) for both the uniform and random chains (with uniformly distributed couplings between $0$ and $1$), 
i.e., the same as the bond-based entropy for the random chain (Fig.~\ref{fig3}). 
The result is expected for the random chain, because the asymptotic SDRG state of such a system is a 
single VB state, in which $S_{\rm loop}=S^1_{\rm VB}=S^2_{\rm loop}$ (and this is the exact result), but it is curious that the
same value obtains for the uniform chain as well. This may also be taken as a flaw of the loop-based entropy (if one wants a
a quantity mimicking the von Neumann entropy as closely as possible), because it is further from 
$S_{\rm vN}$ ($\gamma=1/3$) than the bond-based estimates ($\gamma=4\ln(2)/\pi^2 \approx 0.281$). Nevertheless, it is 
encouraging that the logarithmic divergence is still captured.

In the 2D system, $S_{\rm loop}/\ell$ converges to a finite value with increasing $\ell$ and $L$. Note the rapid convergence as a function of 
the full system size in Fig.~\ref{fig7}. For the largest system ($L=256$) the results are described very well by an area law with an additive 
logarithmic correction; $S_{\text{loop}}=\mu \ell+a\log_2(\ell)+s_0$. The area law prefactor $\mu \approx 0.51$ is, intriguingly, in good agreement 
with $S_{\rm vN}/\ell$ obtained by Kallin {\it et al.} \cite{LADDER} for wide ladder systems [accounting for the $\ln(2)$ included there in the 
definition of $S_{\rm vN}$]. In this case the loop entropy is, thus, a better stand-in for $S_{\rm vN}$ (which is expected to obey the area law) 
than the bond-based quantities.

Note that the presence of long-range N\'eel order in 2D implies the existence of system-spanning loops in the transition graph.\cite{LOOP} 
This explains why $S_{\rm loop}$ is much smaller than $S^2_{\rm VB}$ in Fig.~\ref{fig5} (i.e., because of the large average loop size, the number of loops 
in the transition graph is much smaller than the number of bonds). 


\subsection{N\'eel-ordered 2D amplitude-product state}

We also investigate the scaling behavior of the loop entropy for a 2D amplitude product state, (\ref{singij}) with the expansion coefficients 
given by (\ref{lambdaprod}) with $h(r)=1/r^3$. This corresponds to the asymptotic form of the optimal variational amplitudes for the 2D 
Heisenberg model suggested by previous calculations.\cite{LOOP,LOU} We here do not use fully optimized amplitudes (which show
deviations from the $1/r^3$ for short bonds), because our aim is to investigate a generic N\'eel-ordered state, to compare with the 
results of the specific case of the ground state if the 2D Heisenberg model in Fig.~\ref{fig7}.

\begin{figure}
\includegraphics[width=7.25cm, clip]{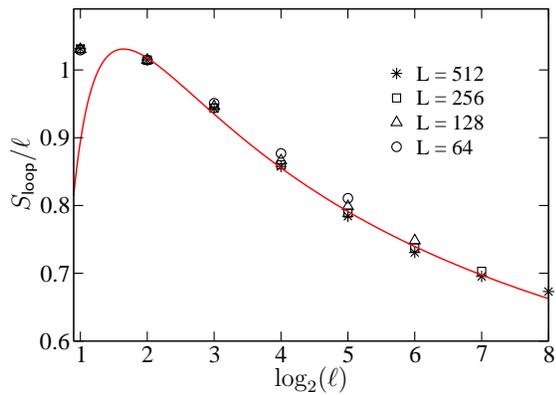}
\caption{
  \label{fig8}
  (Color online).
   Loop entropy of the amplitude-product state with amplitudes decaying with the bond length $r$ as $h(r)=1/r^3$, 
   which is the asymptotic form in the optimal amplitude product state for the 2D Heisenberg model.\cite{LOU} The entropy
   divided by the  system length saturates for large systems, indicating applicability of the area law (with an additive 
   logarithmic correction described by the same form as in Fig.~\ref{fig7}; a fit is indicated by the solid curve).}
\end{figure}

The sublattice magnetization in the amplitude-product state with $h(r)=1/r^3$ for all $r$ is $m_s\approx 0.27$ (extracted from the
system size dependence of $\langle m_s^2\rangle$ for systems of size $L$ up to $256$), somewhat below the known value $m_s\approx 0.307$ for 
the Heisenberg model. The results in Fig.~\ref{fig8} show a behavior very similar to the Heisenberg ground state result in Fig.~\ref{fig7}, 
with a correction to the area law which can be described as an additive logarithm. Because of the lower value of the sublattice magnetization, 
the average loop size is smaller, and, thus, the number of loops in the system (including those shared by the two subsystems) is larger, leading 
to higher overall value of $S_{\rm loop}$ than in Fig.~\ref{fig5}.

\section{Summary and discussion}
\label{summary}

Using the transition graphs characterizing state overlaps in the valence bond basis of $S=1/2$ spins, we have explored two measures of 
bipartite entanglement entropy, $S_{\text{VB}}^2$ and $S_\text{loop}$. The former extends the definition $S_{\text{VB}}^1$ of valence-bond
entropy (Refs.~\onlinecite{VBE1,VBE2}) based on shared valance bonds in the wave function to the transition graph, while the latter 
is based on shared loops (motivated by the loop-gas picture \cite{SUTHER1,SUTHER2} of spin systems). Using an efficient loop algorithm,\cite{LOOP} 
we were able to obtain unbiased QMC results for these quantities in large systems; $2^{13}$ and $2^{16}$ spins in 1D and 2D, respectively. 

In the Heisenberg chain, $S^2_{\text{VB}}$ exhibits a logarithmic divergence, with a prefactor which agrees well with the exact 
value\cite{JACOBSEN} $\gamma=4\ln(2)/\pi^2$ already for small subsystems. In contrast, observing this scaling with the wave-function 
definition $S^1_{\text{VB}}$ \cite{VBE1,LADDER} requires very large systems, due to significant subleading corrections.\cite{CAPPONI} 

For the N\'eel state of the 2D Heisenberg model, both VB entropies violate the area law, exhibiting multiplicative logarithmic corrections.\cite{VBE1} 
We have argued that single-bond definitions typically overestimate the amount of entanglement, because of the over-completeness of the VB basis.  
The loop definition $S_{\rm loop}$ exhibits only an additive size correction to the area law in 2D, in agreement with general expectations 
for standard definitions.\cite{CASINI,RYU,RENYI} 

Important relationships have been established in recent years between the subleading behavior of the entanglement entropy, topological 
order, and quantum-criticality. For instance, the subleading term in a 2D gapped system is a constant determined by the quantum 
dimension of the excitations of the topological phase.\cite{KITAEV,LEVIN} For critical systems in the universality class of $z=2$ 
conformal quantum-critical points, there is a universal additive logarithmic subleading term, which depends only on the shape of 
the subsystem partition and the central charge.\cite{FRADKIN} An additive correction to the area law in the 2D N\'eel state was 
also found for the R\'enyi entropy $S_2$, but the systems were too small to determine its asymptotic form.\cite{RENYI} We have shown 
that the loop entropy $S_{\rm loops}$  captures the essential desired features of an entanglement entropy, scaling logarithmically with 
the subsystem size in 1D and obeying the area law for the 2D N\'eel state. Being relatively easy to calculate, S$_{\rm loops}$ offers 
opportunities to study various aspects of entanglement entropy on large spin lattices in other situations of great interest, e.g., 
at unconventional quantum-critical points.\cite{SENTHIL,JQLOGS} 

In Sec.~\ref{bondentropy} we already commented on the fact that properties such as $S_{\rm VB}^2$ and $S_{\rm loop}$ that are defined using 
specific geometrical properties of the transition graphs (i.e., not following directly from a given operator acting on the spins) in the valence bond 
basis are not necessarily well defined expectation values of some 
hermitian operators. Indeed, it has recently been shown that the entanglement entropies defined in this way are dependent on exactly how a state 
is represented in the overcomplete valence bond basis.\cite{MAMBR} This may suggest that these quantities are ill-defined. However, given a 
hamiltonian $H$, the projector QMC method generates the transition graphs in a unique way, independently of the trial state used (which 
we have also verified explicitly). 

The valence-bond projector technique itself is closely tied to the completely generic ``loop-operator'' representation of the path integral for
a singlet state,\cite{NACHT,LOOP} and therefore the statistical properties of the transition graph loops (including $S_{\rm VB}^2$ and 
$S_{\rm loop}$) are not really tied to the particular QMC scheme, only to the valence-bond basis. The definitions are tied to the hamiltonian, 
in the sense that $H$ generates the transition graphs. We can therefore not, in general, evaluate $S_{\rm VB}^2$ and $S_{\rm loop}$ uniquely just based 
on an arbitrary state, but first need to find the ``parent hamiltonian'' of the state (and furthermore, that parent hamiltonian should have short-range 
interactions only, so that it is unique---support for this generic statement is discussed in Ref.~\onlinecite{VERST}), and use it to generate the 
transition graphs. States defined based on ``unbiased'' bond superpositions
(which includes contributions from all different ways of expressing the state in terms of singlets obeying Marshall's sign rule), such as the 
amplitude-product states (and perhaps generalizations of them including bond correlations) can also be studied, as we have done here in a 2D case 
(and where it is important to note that for such a state with N\'eel order, we obtained results in good agreement with the projected ground state
of the 2D Heisenberg model). In practice, we expect $S_{\rm VB}^2$ and $S_{\rm loop}$ to be useful primarily in QMC studies of ground
states of specific hamiltonians.

In the loop-operator formulation,\cite{NACHT,LOOP} one can think of clusters of spins (defined by transition-graph loops) as forming dynamically in 
imaginary time. The entanglement entropies $S_{\rm VB}^2$ and $S_{\rm loop}$ correspond to entropy measurements averaged over equal-time ``snapshots'' 
of entangled clusters in this time evolution. Exactly how this dynamic aspect relates to the standard wave-function picture of entanglement entropy 
is not clear at present, but the results obtained in this paper suggest that there should be a close relationship between them.
 
\null\vskip-12mm\null
\section*{Acknowledgments}
We would like to thank F. Alet, S. Capponi, C. Chamon, M. Hastings, M. Mambrini, R. Melko, and F. Verstraete
for useful discussions.  YCL acknowledges support from NSC Grant No.~98-2112-M-004-002-MY3 and the 
Condensed Matter Theory Visitors Program at Boston University. AWS was supported by the NSF under 
Grant No.~DMR-0803510 and also acknowledges travel support from the NCTS in Taipei.
 

\end{document}